\documentclass[12pt]{spieman}  
\usepackage{amsmath,amsfonts,amssymb,amstext,aasmacros}
\usepackage{hyperref}
\usepackage{graphicx}
\usepackage{setspace}
\usepackage{tocloft}
\usepackage[T1]{fontenc}
\usepackage[figure,figure*]{hypcap}
\usepackage{deluxetable}
\renewcommand*{\sectionautorefname}{Section} 
\renewcommand*{\subsectionautorefname}{Section} 
\begin{document}
\title{Geometric Distortion Calibration with Photo-lithographic Pinhole Masks for High-Precision Astrometry}

\author[a]{Maxwell Service}
\author[b]{Jessica R. Lu}
\author[a]{Mark Chun}
\author[c]{Ryuji Suzuki}
\author[d]{Matthias Schoeck}
\author[e]{Jenny Atwood}
\author[e]{David Andersen}
\author[e]{Glen Herriot}
\affil[a]{Institute for Astronomy, University of Hawaii, Manoa, \\
640 N. Auhoku Pl. \#209, Hilo, HI 96720 USA}
\affil[b]{Department of Astronomy, University of California, Berkeley \\
Berkeley, CA, 94720-3411, USA}
\affil[c]{National Astronomical Observatory of Japan, 2-21-1 Osawa, Mitaka, Tokyo, 191-8588, Japan}
\affil[d]{Thirty Meter Telescope Observatories, Pasadena, CA, United States}
\affil[e]{National Research Council of Canada - Herzberg, Victoria, BC, V9E 2E7 Canada}

\maketitle
\renewcommand*{\sectionautorefname}{Section} 
\renewcommand*{\subsectionautorefname}{Section} 
\begin{abstract}
Adaptive optics (AO) systems deliver high-resolution images that may be ideal for precisely measuring positions of stars (i.e.~astrometry) if the system has stable and well-calibrated geometric optical distortions. A calibration unit, equipped with back-illuminated pinhole mask, can be utilized to measure instrumental optical distortions. 
AO systems on the largest ground-based telescopes, such as the W.~M.~Keck Observatory and the Thirty Meter Telescope require pinhole positions known to {$\sim$20 nm} to achieve an astrometric precision of 0.001 of a resolution element.
In pursuit of that goal, we characterize a photo-lithographic pinhole mask and explore the systematic errors that result from  different experimental setups. 
We characterized the nonlinear geometric distortion of a simple imaging system using the mask; and we measured 857 nm RMS of optical distortion with a final residual of 39 nm (equivalent to 20 $\mu$as for TMT). 
We use a sixth order bivariate Legendre polynomial to model the optical distortion and allow the reference positions of the individual pinholes to vary. 
The nonlinear deviations in the pinhole pattern with respect to the manufacturing design of a square pattern are 47.2 nm $\pm$ 4.5 nm (random) $\pm$ 10.8 nm (systematic) over an area of 1788 mm$^{2}$.  
These deviations reflect the additional error induced when assuming the pinhole mask is manufactured perfectly square.
We also find that ordered mask distortions are significantly more difficult to characterize than random mask distortions as the ordered distortions can alias into optical camera distortion. 
Future design simulations for astrometric calibration units should include ordered mask distortions.
We conclude that photo-lithographic pinhole masks are $>10$ times better than the pinhole masks deployed in first generation AO systems and are sufficient to meet the distortion calibration requirements for the upcoming thirty meter class telescopes.
\end{abstract}
\section{Introduction}
Precise astrometric measurements from have enabled scientific results across a variety of fields. This includes studies of stellar clusters that utilized proper motions to identify members and study dynamical structure \cite{Massari:2015,Hosek:2015,Platais:2018}, orbital measurements of nearby binaries of low mass stars and exoplanets \cite{Rodet:2018,Best:2017,Wertz:2017,Konopacky:2016,Dupuy:2018}, and orbital measurements of the stars at the Galactic Center \cite{Ghez:2008,Lu:2009}.
These science cases require sub-milliarcsecond precision, which is typically $10-100\times$ smaller than the intrinsic geometric distortion in the instrument so, the distortion must be measured and corrected for each instrument.  

The best current distortion calibrations for astronomical instruments use observations of crowded stellar fields to simultaneously solve for the static optical distortions of the imaging system and the intrinsic on-sky positions of each star\cite{Anderson:2003,GAIADR2:2018}.
This technique, often referred to as the self-calibration method, requires translating and rotating the pointing of the telescope many times in order to move the stellar cluster across the field of view and constrain all possible distortion modes \cite{Anderson:2003}. 
Self-calibration can also be applied to a calibration unit using artificial sources as long as the astrometric reference positions can be rotated and translated.
A self calibration method that included time variation was adopted for Gaia Data Release 2 (DR2), which delivered an absolute astrometric calibration with uncertainties of $<$0.04 mas for the brightest sources \cite{GAIADR2:2018}.
The high precision of the self-calibration method results from the quantity and diversity in the data, which cannot easily be replicated for all astronomical imaging systems as the observing time commitment is too large.
Instead, most ground-based instruments adopt an external set of calibrated stellar positions (generally HST or Gaia positions) as distortion free and model the distortion as the difference between measured positions and the external catalog \cite{Yelda:2010,Service:2016,Maire:2016, Massari:2016}.
This approach still requires observing time to measure the stellar field, but it is greatly reduced from the requirements of a full self-calibration.

The requirement of using on sky observations for distortion calibration could be eliminated by using an internal astrometric flat field to measure the distortion.
A natural candidate for this astrometric flat field is a pinhole mask with a regular grid of holes at precisely known positions.
This is not a new idea; however, previous attempts have failed to match the accuracy that can be reached using images of star fields.
For example, the first distortion maps for the Near Infrared Camera (NIRC2) instrument at the W.M Keck Observatory were measured using pinhole masks \cite{NIRC2:2001,PBC:2007}; but, these solutions had residual distortion $>$2 mas as compared to 1 mas residuals from  solutions derived with globular clusters \cite{Yelda:2010,Service:2016}. 

More recently, the distortion of the Gemini Planet Imager was calibrated using a combination of a pinhole mask with unknown hole positions combined with the self-calibration method \cite{Konopacky:2014}. 
They achieved a distortion residual of 0.56 mas over a 2.67$^{\prime\prime}$ $\times$ 2.73$^{\prime\prime}$ field of view.
The improvement in distortion measurement over previous attempts was due to the use of the self-calibration method, not improvement in the manufacturing precision of the mask. 
This approach requires a translation and rotation stage in the calibration unit, which is not always feasible; however it's worth noting that even large manufacturing errors can be mitigated with this method.
  
The quality of available pinhole masks determines the optimal design for a given instrument calibration unit.
If the residual pattern errors (mask distortion) in the mask are less than the required distortion calibration precision, one can adopt the manufactured pattern as the "distortion free" reference and simply accept that there will be residual distortion in the final solution due to the ignored error.
This can be achieved either by accurately manufacturing the mask pattern or by measuring the mask distortion before it is installed in the calibration unit.  
An accurate known astrometric flat field greatly relaxes the functional requirements for the system, as a rotation and translation stage is no longer required. 
It would also greatly reduce the amount of observing time required to measure the distortion, which is particularly important if the calibration has to be repeated to account for variations in the instrument.
One drawback is that distortion solutions measured using an internal mask would be blind to optical distortion in the telescope itself, however the distortion in most high resolution AO fed astronomical cases is dominated by the distortion intrinsic to the instrument.
For example, comparison of models and the measured on-sky distortion solution for the GeMS system demonstrated that the optical distortion is dominated by the AO system.\cite{Patti:2019}.
This is consistent with the distortion estimates based on the Zemax optical prescription for NIRC2 which shows that the AO relay contributes 10000 times more distortion than the telescope over a 14" $\times$ 14" field of view.

The goal of this work is to understand the error contribution to a final distortion calibration due to manufacturing errors in a reference pinhole mask and the distortion measurements errors induced by the calibration procedure, as this directly informs the design of future calibration units.
We specifically focus on potential applications to the first-light Narrow-Field Infrared Adaptive Optics System (NFIRAOS) at the Thirty Meter Telescope (TMT) \cite{Herriot:2014} and the narrow field mode of the infrared camera (NIRC2) at the W.~M.~Keck Observatory.
Using a prototype pinhole mask provided by NRC Herzberg, we quantify the deviation of the pattern of pinholes with respect to a perfect square pattern.  
For TMT, the final requirement for the total astrometric error budget is less than 10 $\mu$as, which means that the residual distortion must be much smaller.
This converts to a physical size of 20 nm at the telescope focal plane inside NFIRAOS. 
Similarly for Keck NIRC2, the residual distortion due to manufacturing errors must be significantly smaller than current distortion calibration residual of 270 nm or 1.0 mas \cite{Service:2016}.  
If the mask pattern is accurate to the 20 nm level, then a calibration unit for either TMT NFIRAOS or Keck NIRC2 could be completely static.  

We present results from both a laboratory experiment and simulations that show that the required astrometric calibration precision can be achieved using current pinhole masks and small dithers and rotations. \S\ref{sec:experiment} describes the experimental setup, \S\ref{sec:obs} describes the observations obtained, and \S\ref{sec:analysis} describes the analysis procedures employed to extract pinhole positions and fit distortions. In \S\ref{sec:simulation}, we present a set of simulations that reproduce the experimental results and are extended to explore the impact of different dithering and rotation schemes during calibration and to show the sensitivity to manufacturing errors in the pinhole mask. Finally, in \S\ref{sec:discuss}, we discuss how our results effect the design of future distortion calibration units.

\section{Experimental Setup}
\label{sec:experiment}
\begin{figure}
    \centering
    \includegraphics[width=0.5\textwidth]{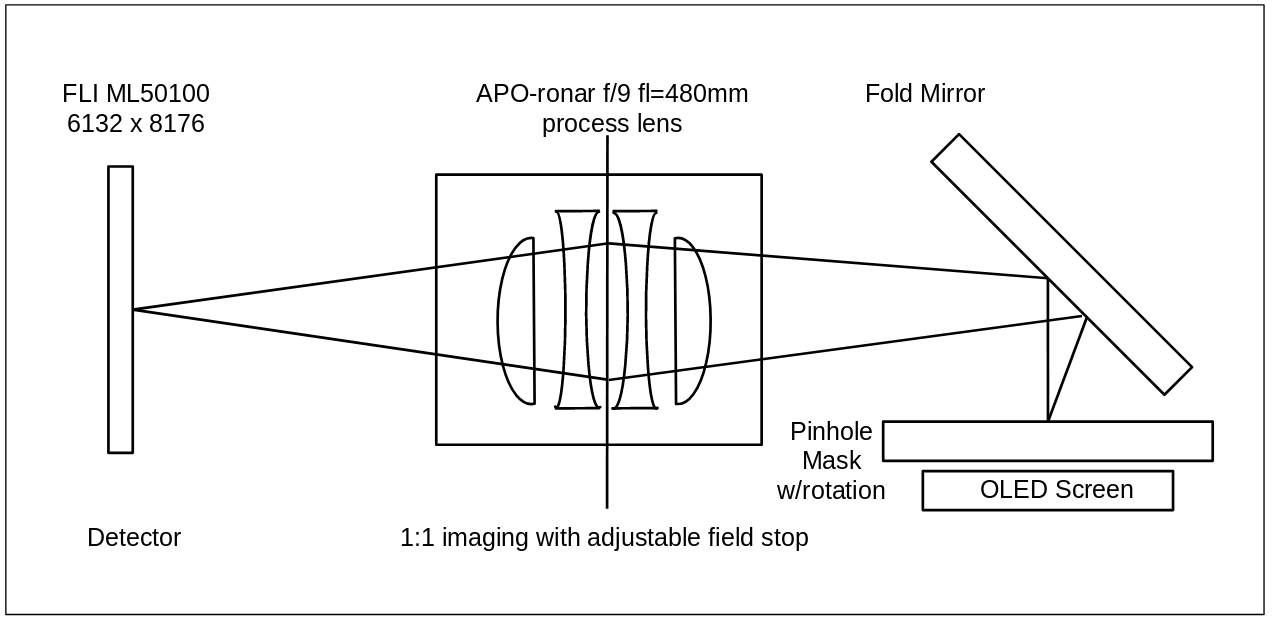}
    \caption{Diagram showing the lab setup used to image the pinhole mask.}
    \label{fig:lab_schematic}
\end{figure}
A lab experiment was designed to measure the accuracy of the pinhole mask hole positions and sizes.  
The lab setup consists of a light source (OLED cell phone screen) illuminating the pinhole mask, which is then imaged using a low-distortion field lens on to a large-format optical camera as shown in Figure \ref{fig:lab_schematic}.   
The pinhole mask was mounted parallel to the lab bench with a simple three point mount on a rotation stage, and a flat mirror was used to fold the beam into the lens.
The lens is an Apo-Ronar process lens with a 480 mm focal length designed for low distortion 1:1 imaging when used at $f/9$. 
We used a Finger Lake Imaging (FLI) CCD camera (model ML50100) with a 8176 $\times$ 6132-element detector with 6 $\mu$m pixels (61.2 mm diagonal) and a quantum efficiency (QE) $>$ 30 $\%$ from 360 nm to 800 with a peak of 60\% at 540 nm.

\begin{figure}[t]
\centering
\includegraphics[width=.8\textwidth, trim={0 0 0 0}, clip]{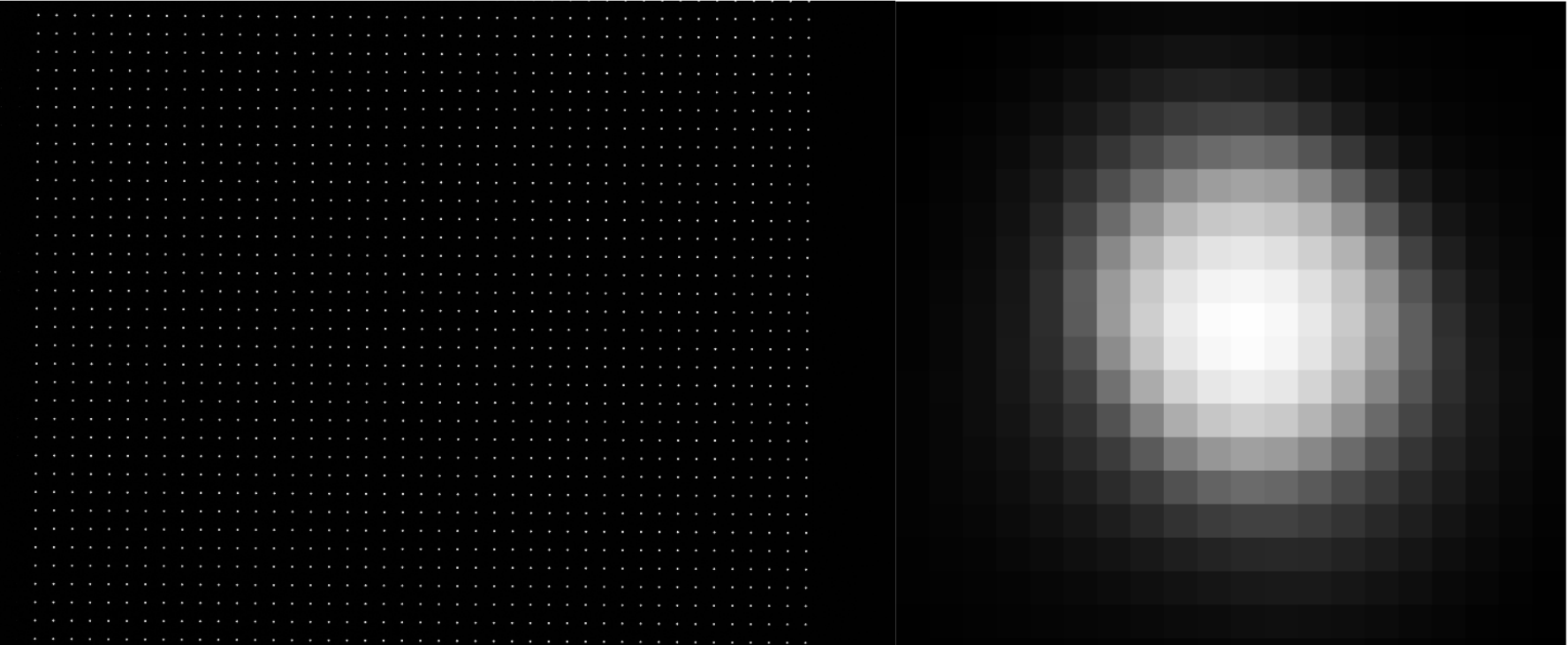}
\caption{{\em Left:} Example FLI CCD image of the 56 $\mu$m-sized pinholes which were used to measure the optical distortion in this setup. The size of the image on the left is 46 mm $\times$ 35 mm  {\em Right:} The spot shape or PSF empirically determined for this image using \textit{StarFinder} as discussed in section 3. The PSF cutout is 0.12 mm on each side and has a linear intensity stretch. \label{fig:ex_im}}
\end{figure}

The pinhole mask is a prototype obtained from NRC-Herzberg and the TMT NFIRAOS project produced by Advance Reproduction using photo-lithographic techniques by Advance Reproduction in chrome on a fused silica wafer.   
The mask was made on a 125 mm diameter Quartz wafer with a chrome on nickel coating that has an optical depth of 3.
In addition there is a coating of {\it Advance Polyguard}, which is a thin transparent film with anti-wetting, anti-stiction and anti-migration properties.
The mask has 4 different pinhole  diameters (12, 24, 56 and 120 $\mu$m) situated in a 86$\times$86 square pattern with 1 mm spacing between each pinhole and an expected tolerance on the diameter of $<$ 0.3 $\mu$m.  
When imaging the small (12$\mu$m and 24 $\mu$m) pinholes, we found that there was a systematic position measurement error of $\sim$ 100 nm, which we attribute to small scale detector defects.
As a result we used the 56 micron pinholes for this work; an example image is shown in Figure \ref{fig:ex_im}
As this experiment is done using visible light, there is background transmission of $\sim$ 1\% between the pinholes.    

There were 2 different experiments conducted with this setup: (1) obtain multiple images of the pinhole mask without dithering to analyze the measurement precision and stability over time, and (2) dither the pinhole mask to derive the optical distortion and the pinhole positions simultaneously without assuming the pinhole pattern is perfectly square (self-calibration method).  
For the second experiment we consider 6 mask positions that are spaced equidistantly around a unit circle.
The rotation stage is displaced from the center of the detector by $\sim$3.5mm in both axes, so rotating the mask serves to both rotate and translate the pattern with respect to the detector.
We only consider positions from pinholes that are in at least 3 of the 6 locations, which eliminates 61 pinholes leaving a total of 1788 pinholes sources used.

For images at a single mask position, both optical distortions and irregular pinhole positions will manifest as deviations from a regular grid and we cannot distinguish between the two sources. 
However, the optical distortions are static with respect to the camera; thus, moving the mask with respect to the camera allow us to separate optical distortions from pinhole irregularities. 
For the final analysis, data is taken after rotating the pinhole grid to separate mask distortion and camera (optical) distortion. 
\section{Observations and Data}
\label{sec:obs}
\begin{figure}
    \centering
    \includegraphics[width=0.5\textwidth]{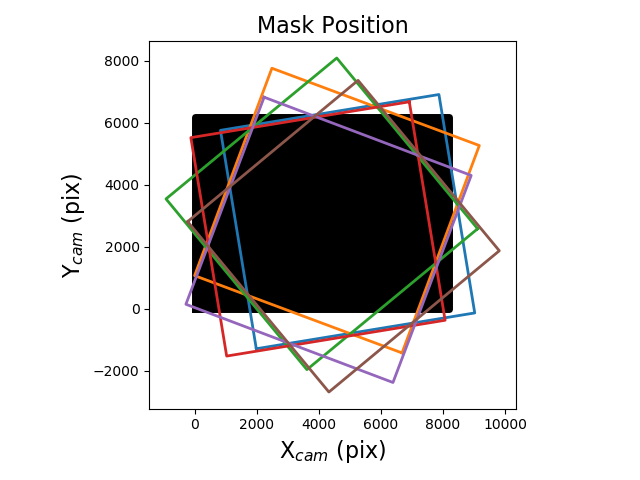}
    \caption{Positions of the pinhole pattern with respect to the camera detector.  The shaded black region shows the detector and colored squares show the outline of the pinhole pattern for each of the 6 locations the mask was observed. The images were taken with a plate scale of 6 microns per pixel and the axis of rotation of the mask is offset from the center of the detector by 3.5 mm.}
    \label{fig:pos}
\end{figure}

The observations are summarized in Table \ref{tab:data} and consists of 6 stacks of images taken with the mask rotated to a different position for each stack as shown in Figure \ref{fig:pos}.  
Each rotation position was imaged once with a total of 100$\times$0.3 s exposures.  
The axis of rotation is offset from the center of the detector by approximately 3.5 mm in both axis. 

Raw images were dark subtracted and flat field corrected. The flat field was measured using images taken with the illumination on and pinhole mask removed from the system.   
Each image was then run through a source extraction procedure to identify the pinhole images and measure their positions and fluxes. 
The source extraction is performed using the point spread function (PSF) fitting routine, {\em StarFinder} \cite{Diolaiti:2000} with a PSF box size of 20 pixels. 
The PSF is determined empirically from the data and is the average of all sources in a 2000 pixel box centered on the detector.  
The output of {\em StarFinder} is a catalog of pinhole positions and fluxes in raw detector coordinates for each image. These output catalogs will be the input for the averaging and model fitting in the next sections.

\section{Experimental Analysis}
\label{sec:analysis}

\subsection{Stability}
\begin{figure}[h]
\centering
\includegraphics[width=\textwidth]{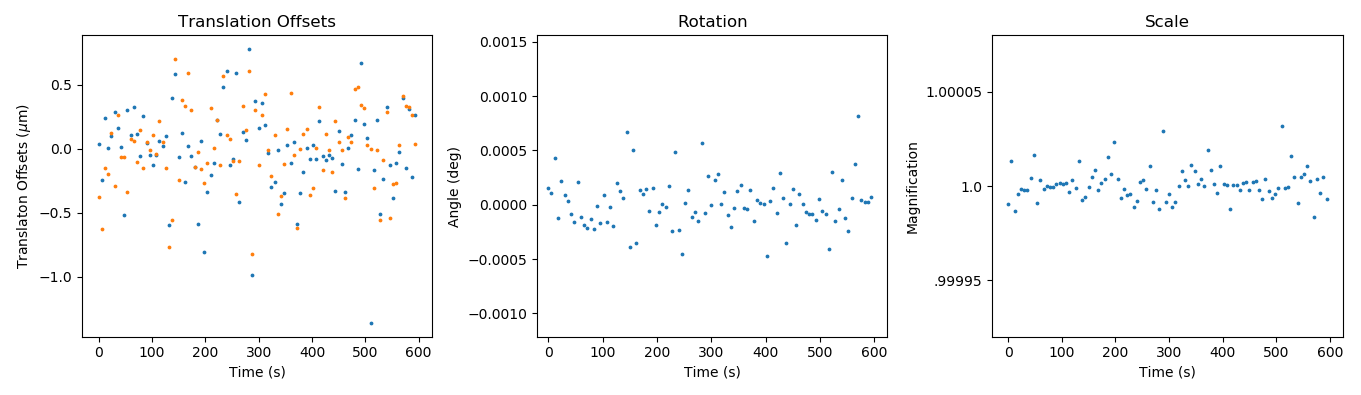}
\caption{Variation in computed 4-parameter transformation coefficients for each of the catalogs derived from the 100 exposures.  The translation and rotation drift are consistent with motion in the rotation stage, while the variation in the scale is much smaller and is likely due to instabilities in the optics.  The pixel size is 6 $\mu m$ and a rotation of 0.0001 degrees corresponds to a tangential motion of 110 nm at the edge of the rotation stage. \label{fig:stability}}
\end{figure}

\label{secp:precision}
\begin{figure}[b]
\centering
\includegraphics[width=0.5\textwidth]{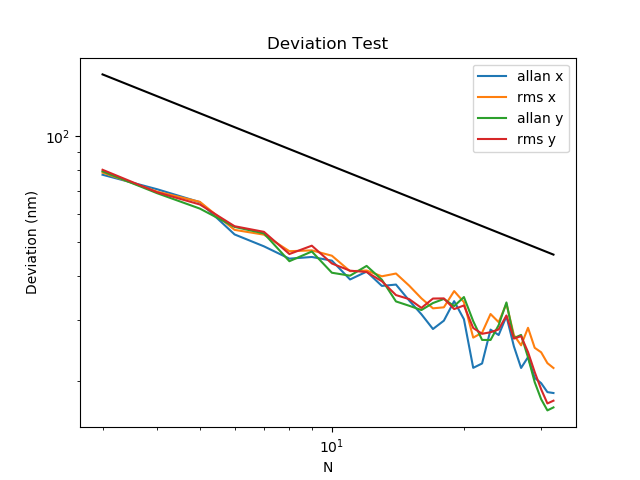}
\caption{\label{fig:deviation} Allan deviation and RMS deviation for a sequence of 100 exposures taken at the same mask position. The black line shows the expected N$^{0.5}$ behavior. 
As expected, the deviations fall as $N^{-0.5}$ (black line).  
This behavior is only seen after a 4 parameter fit is used to reference the catalogs, due to the instability in the lab setup.}
\end{figure}
Before trying to separate true optical distortion from possible errors in the mask, we must first understand the stability of the setup at a single mask location.  
We analyze the variation of the position measurements in a stack of catalogs derived from images taken with the mask at a fixed location.  
This analysis does not constrain the optical distortion in the system; it only estimates the experimental stability.
For this test we use a stack 100 source catalogs with 1493 detected images of the 56 micron pinholes.    
We test the stability by verifying that the scatter in the measured mean position decreases as a function of the number of frames used in the mean. 
The stack of catalogs are split into N groups with M catalogs, and the mean position is computed for each group.  
Then the RMS deviation and the Allan deviation is calculated for the N groups of position measurements. 
If the scatter in the position measurements is due to random errors, than the scatter should be proportional to $N^{-0.5}$.   
When this test is performed with only the mean translations eliminated, the RMS deviation does not decrease as a function of N, which implies that there is significant variation in the scale and rotation of the images as seen in Figure \ref{fig:stability}. 
When we increase the complexity of the transformation to include the scale, rotation and translations (4 parameters) than the scatter in the measured positions decrease as $N^{-0.5}$.  

The four parameter fits are performed iteratively, by first averaging all the catalogs with only translation removed to create the first set of reference coordinates, and then fitting a new four parameter transformation between each catalog and these reference coordinates. 
The new transformed coordinates are then averaged to create a new reference and the fitting procedure is repeated a final time.  
The resulting corrected positions show the expected behavior in the scatter as a function of the number of measurements averaged (Figure \ref{fig:deviation}).  
Given these results, we remove the linear parameters and then  average over the coordinates in each stack and take the error on the mean as the measurement uncertainty.  We produce one stack-catalog per dither and rotation position and use the stack-catalogs going forward. 

\begin{figure}
\centering
\includegraphics[width=0.3\textwidth]{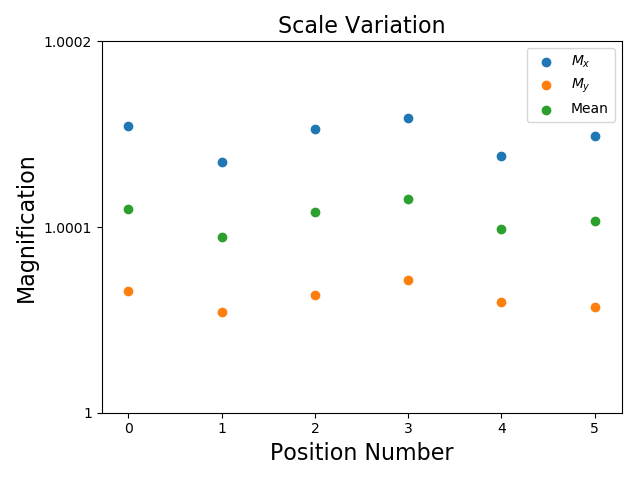}
\caption{The variation in the magnification for each of the 6 positions used to measure the optical distortion of the system as summarized in Table \ref{tab:data}.  The linear scales $M_{x}$ and $M_{y}$ are computed as $(c_{1}^{2} + c_{2}^{2})^{0.5}$ where the $c$ coefficients are defined in equation \ref{eq:linear}. 
A change of 0.0005 in the scale corresponds to a $\Delta_{z}$ along the optical axis of 240 microns.  These variations are eliminated as part of the complete distortion model, however, this means that this analysis is not sensitive to variation in these terms that is intrinsic to the mask.  The dominant change is in the global scale with only small variations of the skew (ratio between $M_{x}$ and $M_{y}$).  \label{fig:stack_stability}}
\end{figure}
The averaged positions of the 100 catalogs at each position of the mask will be used to measure the optical distortion of the setup (camera distortion) and the position of the pinholes on the mask (mask distortion) as described in the following sections.  

\begin{figure}[h]
\centering
\includegraphics[width=0.9\textwidth]{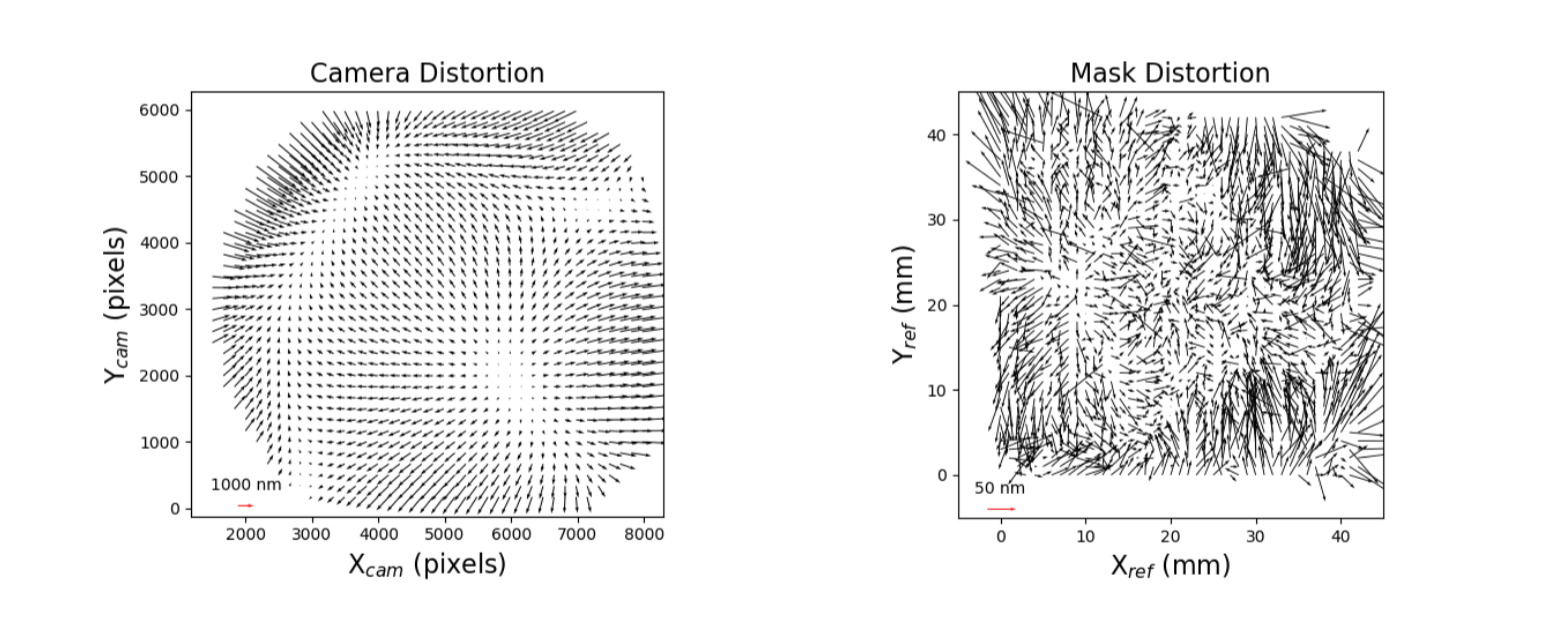}

\caption{Above are the best fit models for the camera and mask distortion.  The camera distortion is plotted as a function of the camera pixel (6 $\mu$m pitch) and has an RMS size of 1005 nm and 711 nm in x and y respectively. For plotting purposes, the camera distortion is sampled every 166.7 pixels (1 mm).  The mask distortion is plotted in physical mask coordinates, where the pinhole pattern extends 43$\times$43 mm and has an RMS size of 35.6 and 57.9 nm in X and Y respectively. Note that the scale of the arrows is different in the two panels and the camera distortion is $\sim$16$\times$ larger.
\label{fig:measured_dist}}
\end{figure}
\begin{figure}[h]
\centering
\includegraphics[width=0.9\textwidth]{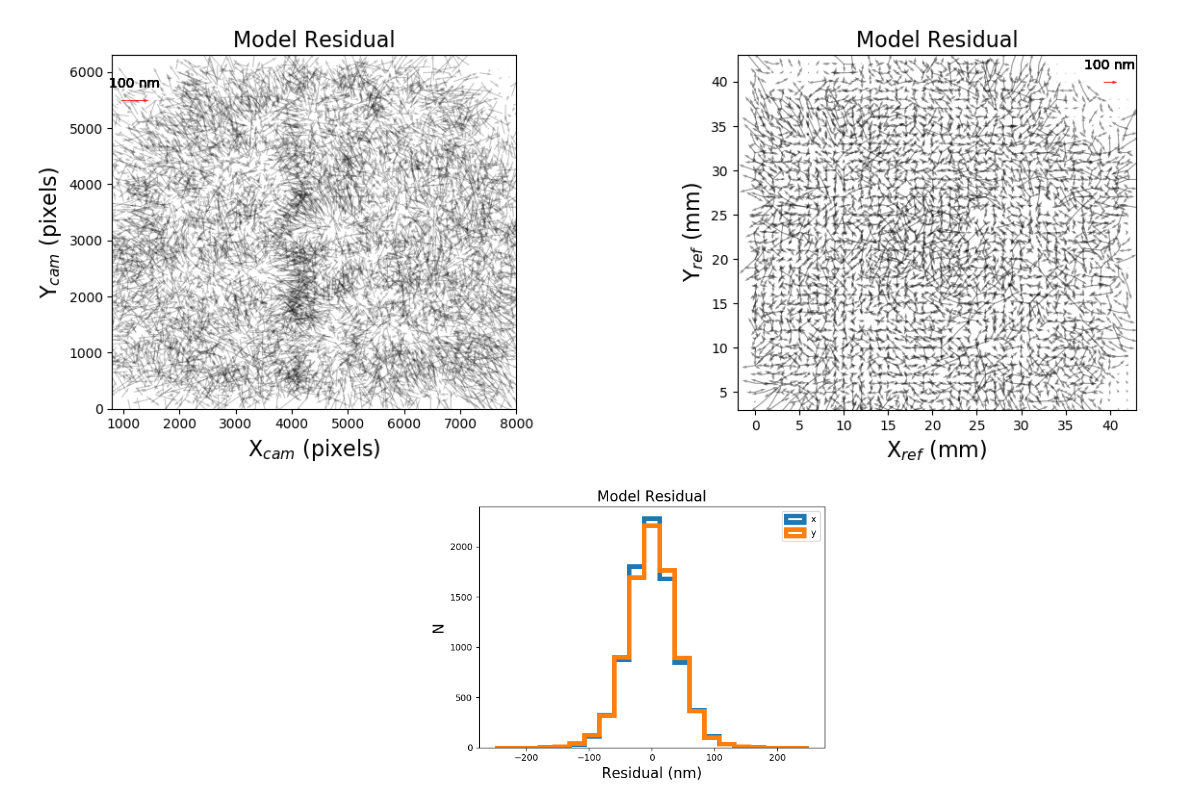}
\caption{The residuals of the complete model that includes camera distortion and mask distortion.  The data have been trimmed to only include pinholes measured in at least 3 of the 6 mask positions.  \textit{Left}:Residuals plotted with respect to the camera pixels.  The spatially correlated residuals are due to high order distortion that cannot be fit by our distortion model.\textit{Right}:Residuals plotted with respect to the pinhole mask. \textit{Bottom}: Histogram of the residuals to the complete model, the RMS scatter is 39.5 nm and 38.5 nm in X and Y respectively.   
\label{fig:residual_dist}}
\end{figure}

\subsection{Distortion Models}
\label{sec:dist_model}

In order to constrain the camera distortion, the mask distortion, and control for experimental instabilities, we fit a 3-component model to the stack-catalogs from all the dither and rotation positions simultaneously. 
he dither position coordinates are listed in the last 3 columns of Table \ref{tab:data}.
 We define two coordinate systems, 
$$x_c, y_c  \;\;\;\textrm{camera coordinates}$$
$$x_m, y_m  \;\;\;\textrm{mask coordinates}$$
where we adopt a uniform square grid of pinholes with a pitch of 166.7 camera pixels as our mask coordinate frame. The camera coordinates are an idealized distortion free coordinate frame that is defined with respect to the detector pixels.

The three components of the model include the following. 
First, there is 
\begin{equation}
\begin{split}
    x_c = D_{x,camera}(x'_c, y'_c)\\
    y_c = D_{y,camera}(x'_c, y'_c)
    \end{split}
\end{equation}
where
\begin{equation}
D_{camera}(d, x'_c, y'_c) = \sum_{i=0}^{R}\sum_{j=0}^{R}d_{ij}L_{i}(x_c)L_{j}(y_c)
\label{eq:distortion}
\end{equation}
which defines is the transformation from distorted camera coordinates($x'_c, y'_c$) to distortion free coordinates ($x_c, y_c$). Here $L_{i}$ is the $i^{\textrm{th}}$ normalized Legendre Polynomial and $R$ is the order of the distortion model which is 6 in this work. When this function is evaluated, the arguments are normalized to lie on the interval from -1 to 1. This model does not include the linear parameters ($i+j < 1$) as those are included in the second component of this model.
Second, there is  
\begin{equation}
\begin{split}
x_c = l(x_m, y_m) = c_{0} + c_{1}x_m + c_{2}y_m \\y_c = l(x_m, y_m) = c_{3} + c_{4}x_m + c_{5}y_m
\end{split}
\label{eq:linear}
\end{equation}
 which defines a linear transformation to go from the mask coordinate frame ($x_m$,$y_m$) to distortion corrected positions in the camera coordinate frame.  As the mask is rotated on the detector, these linear parameters are different for every stacked catalog.
Third, there is the mask distortion,
\begin{equation}
\begin{split}
D_{x,mask}(x_{square}) = x_{m} + \delta_x \\
D_{y,mask}(y_{square}) = y_{m} + \delta_y
\end{split}
\end{equation}
where $x_{m}$ ($y_{m}$) are the original uniform grid of pinhole positions, and  $\delta_x$ ($\delta_y$) are the differences between the uniform grid and the true pinhole positions. 
The correct pinhole defined as 
\begin{equation}
\begin{split}
    x_{pin} = D_{x,mask}(x_{m}) \\
    y_{pin} = D_{y,mask}(y_{m}) 
    \end{split}
\end{equation}
where $x_{pin}$ and $y_{pin}$ are the true pinhole positions on the mask.
To fit this model, we must specify all three model components.  
This is done in an iterative fashion, where we start by assuming $\delta_x$ and $\delta_y$ are zero and fitting the first two model components using a Levenberg-Marquardt to simultaneously fit for the $d$ and $c$ coefficients.
To do this we use the measured coordinates from each stack catalog as $x_c'$ and $y_c'$ and minimize $\Delta$ 
\begin{equation}
\begin{split}
\Delta_x =  \sum_{n=0}^{N-1}[l(c_{n}, x_{pin}, y_{pin}) - (D_{\textrm{x,camera}}(d,x_{c,n}', y_{c,n}'))]^2
\\
\Delta_y =  \sum_{n=0}^{N-1}[l(c_{n}, x_{pin}, y_{pin}) - (D_{\textrm{y,camera}}(d,x_{c,n}', y_{c,n}'))]^2
\label{eq:minimization}
\end{split}
\end{equation}
where N is the total number of catalogs (6) and $n$ denotes the $n^{th}$ catalog.
Note that there are separate linear parameters ($c$) for each of the $n$ catalog, which gives a total of 134 free parameters.
After the camera distortion and linear parameters have been fit, we update the model of the mask distortion.
This is done by applying the current best model of the camera distortion to the measured positions and inverting the linear equations (equation $\ref{eq:linear}$) to transform those distortion corrected positions into the mask coordinate frame.
The mask distortion is then corrected as 
\begin{equation}
    \begin{split}
        \delta_x = \bar x_m' - x_{m} \\
        \delta_y = \bar y_m' - y_{m}
    \end{split}
\end{equation}
where $\bar x_m'$ and $\bar y_m'$ are the average of the distortion corrected measured positions in the mask reference frame.
After updating the model for the pinhole positions, the camera distortion model and the linear parameters must be fit again.
This fitting procedure is then repeated with the new values for $\delta$ for 4 iterations, when the change in the mask distortion model is $<$ 20 nm.
Note that the linear transformations in this model mean that this analysis is entirely blind to linear modes of camera and mask distortion.
The linear transformations are required to eliminate the variation seen in the optical system between measurements of the different pinhole mask positions as shown in Figure \ref{fig:stack_stability}.
\begin{deluxetable}{cccccccccc}
 \tabletypesize{\footnotesize}
\tablecolumns{8} 
\tablewidth{0pt}
 \tablecaption{Data Summary
 \label{tab:data}}
 \tablehead{
 \colhead{Date}
 &\colhead{Position}
 &\colhead{N$_{sources}$}
 &\colhead{N$_{exp}$}
 &\colhead{T$_{exp}$ (s)}
 &\colhead{$\sigma _{x}$(nm)}
 &\colhead{$\sigma _{y}$ (nm)}
&\colhead{Angle (deg)}
&\colhead{$\Delta_{x}$ (mm)}
&\colhead{$\Delta_{y}$ (mm)}
}
\startdata
01-05-2019 &0 & 1353 &100 & 0.3 & 9.9 & 10.1 &10&0 &0\\
01-05-2019 &1 & 1373&100 & 0.3 & 11.8 & 12.5 &70&4&1.6\\
01-05-2019 &2 & 1346&100 & 0.3 & 7.6  & 8.0 &130&4.6&5.9\\
01-05-2019 &3 & 1372&100 & 0.3 & 7.6  & 8.1 &190&1.3&8.5\\
01-05-2019 &4 & 1337&100 & 0.3 &8.0  &9.7 &250&-2.7&6.9\\
01-05-2019 &5 & 1312&100 & 0.3 &8.9  &8.9 &310&-3.3&2.7\\
\enddata
\end{deluxetable}
  
\subsection{Estimated Distortion Precision}
\label{sec:dist_results}

Our best fit has a 5-$\sigma$ clipped RMS residual of 39.5 in X and 38.5 nm in Y. 
These residuals are a combination of the positional measurement error and the residual optical distortion with order $>$ 6.
We estimate the residual high order distortion as the total fit residual subtracted by the position measurement errors in quadrature (37.8 nm).
Table 2 summarizes the contributions to the total position displacements in the system, which are the optical camera distortion, the mask distortion, the position measurement error and the residual optical distortion.
Figure \ref{fig:measured_dist} shows the best fit models for the camera and mask distortion.
The size of the camera distortion is consistent with the manufacturer's specification of less than 0.01\% distortion. 
Mask distortion is estimated to be 47.2 nm RMS, which would limit the accuracy of distortion measurements using this mask if it is not pre-calibrated.  
This sets the expected floor for distortion measurements carried out using other similar masks assuming the new mask was not independently calibrated.
The spatial coherence in the fit residuals seen in Figure \ref{fig:residual_dist} suggests that the uncorrected high order optical distortion is a significant contributor to the fit residual.
High spatial frequency defects in the detector could also be contributing to the remaining residual. 

It is worth noting that environmental instability in the system could contribute to the residual term that we attribute to residual high order distortion.  
For example, if there is a changing temperature gradient on the pinhole mask during the observations, it would alter the mask distortion pattern between different mask positions.
As we assume a static mask distortion, this change would increase the residual in the fit.
This is generally true for any instability which changes the mask or camera distortion in a non-linear fashion during the experiment.
Note that if the temperature of the entire mask changes then there would be no change in the mask distortion, as the expansion only effects the linear terms in the mask distortion which are eliminated in our model.
\begin{deluxetable}{cccc}
 \tabletypesize{\footnotesize}
\tablecolumns{8} 
\tablewidth{0pt}
\tablecaption{Deviation Budget}
\label{tab:final}
\tablehead{
\colhead{Source}&
\colhead{Size (nm)}&
\colhead{Size TMT ($\mu$as)}
}
\startdata
Total Nonlinear Deviations &858&429\\
\hline
Optical Distortion $\mathcal{O}$(2-6)&856&300\\
Pinhole Mask Distortion &47.2 $\pm$4.5 $\pm$11& 23.6\\
Measurement Precision&9.2&4.6\\
Uncorrected high order distortion $\mathcal{O}$(>6) &37.8&18.9\\
\enddata
\end{deluxetable}

\section{Simulations}
\label{sec:simulation}

In order to evaluate the measurement error in the mask distortion we must understand how effective the self-calibration method was in this case.
The aim of self-calibration is to correctly identify the deviations due to the mask distortion and the deviations due to the camera distortion.
Self calibration approaches have been widely implemented for non-astronomical imaging systems and have been found to have significant degeneracy between the fit parameters.
For example, Strum et. al. studied the degenerate solutions in a model of the 5 intrinsic camera parameters and discovered a certain class of camera moves where there are multiple solutions for the focal length.  
The same problem applies for the more complicated models of optical distortion, where there are multiple combinations of mask distortion and optical distortion which fit the data equally well\cite{Strum:2000}. 
This is a general problem for camera self calibration techniques, that has been explored in the specific case of radial distortion \cite{Brito:2013,Wu:2014}.
As our model includes both radial and tangential optical distortion their results only reflect a subset of the degenerate cases that could be present in our system.
Brito et. al. note that pure translation in the XY plane fails to accurately recover the radial distortion\cite{Brito:2013}, which is consistent with the simulated results in this section and others have found a number of other possible motions with more than one valid distortion solution.\cite{Wu:2014}
Another important case those authors notes is that rotation about the optical axis is degenerate, however, rotation about any other point is not. 
As the cases in the literature differ substantially from our setup, and the size of the error depends on the strength of the distortion, we choose to simulate our optical setup to estimate the systematic errors in the measured mask distortion.  

The simulations presented here use two distinct starting points (1) a realistic set of inputs which match the best fit measurements for the mask and camera distortion from the real data (Figure \ref{fig:2d_sim_best_recovery_error}) and (2) a worst case scenario where the optical distortion is set to zero while the total distortion in the system is applied as mask distortion (Figure \ref{fig:sim_all_dist}).
The second set of inputs is not realistic, however it maximizes the error in recovery because if a given deviation can be fit as either mask distortion or camera distortion it will be measured as camera distortion.
We use the first simulation to estimate the random errors in the mask distortion and the second simulation to estimate the systematic error due to mask deviations being incorrectly characterized as optical distortion.

\subsection{Accuracy of the Measured Mask Distortion}
\label{sec:sim_errors}

\begin{figure}
    \centering
    \includegraphics[width=0.9\textwidth]{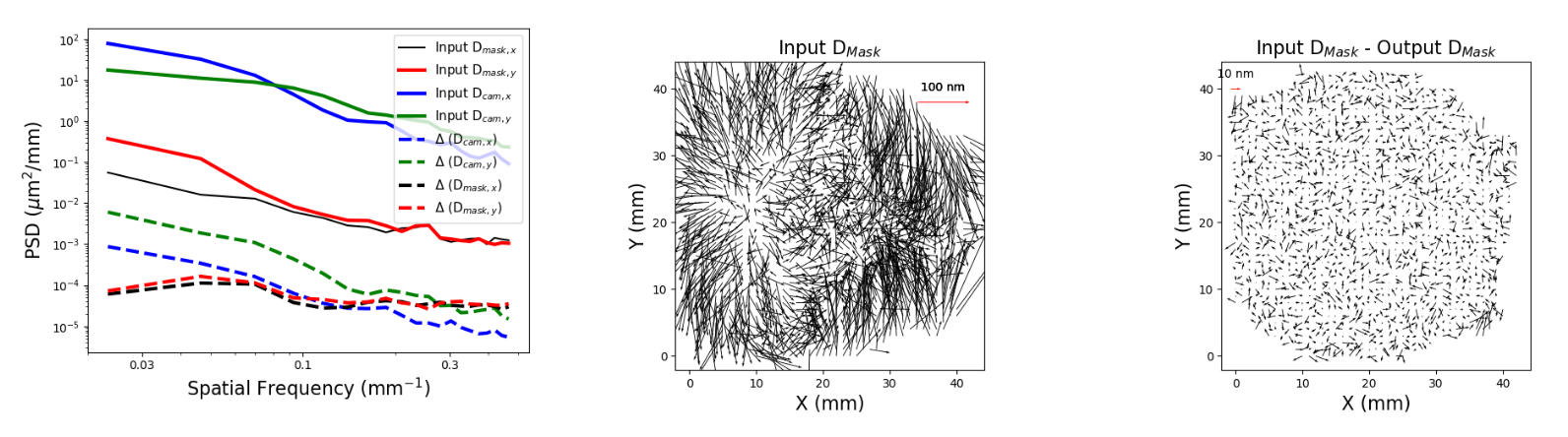}
    \caption{Results from the simulation where the inputs match the optical system in the lab.  The best fit solutions are used as the input mask and camera distortion and the same reference mask positions are for the 6 simulated data sets. \textit{Left}:Power spectrum distribution for the input mask distortion and the residual between the input and measured mask distortion. The $\Delta$ quantities are the difference between the input distortion and the distortion recovered by the simulation for the mask and camera distortion respectively.  Both are evaluated from a square grid of points with 1 mm spacing.  \textit{Center}:Input mask distortion for these simulation, this matches the measured mask distortion measured on the lab setup with a size of 46.8 nm RMS. \textit{Right}:Difference between the input and recovered mask distortion. This difference has an RMS size of 4.5 nm. Note that this simulation includes random measurement errors per catalog as reported in Table \ref{tab:data}. }
    \label{fig:2d_sim_best_recovery_error}
\end{figure}

Using those two sets of inputs, we first create a simulation to replicate the actual lab experiment using simulated catalogs instead of the real data to estimate the error in the mask distortion measurement.
To create the simulated observations the best fit mask distortion is applied to a perfectly square grid of points to create a set of simulated mask coordinates.
We then use the same linear parameters from the best fit to transform the simulated mask coordinates into the camera coordinate frame, apply the best fit model of the distortion as a function of the camera coordinates and add Gaussian noise consistent with the measurement errors in Table \ref{tab:data}.  
This gives 6 simulated measured catalogs of the pinhole mask that match the 6 real observations in the lab.
The simulated data is then fit with the same model described in Section \ref{sec:analysis}, and the mask deviation is recovered with an accuracy  of 4.5 nm RMS as shown in Figure \ref{fig:2d_sim_best_recovery_error}.
We adopt this value as the random component of the error in the measurement of the mask distortion.

\begin{figure}
    \centering
    \includegraphics[width=0.9\textwidth]{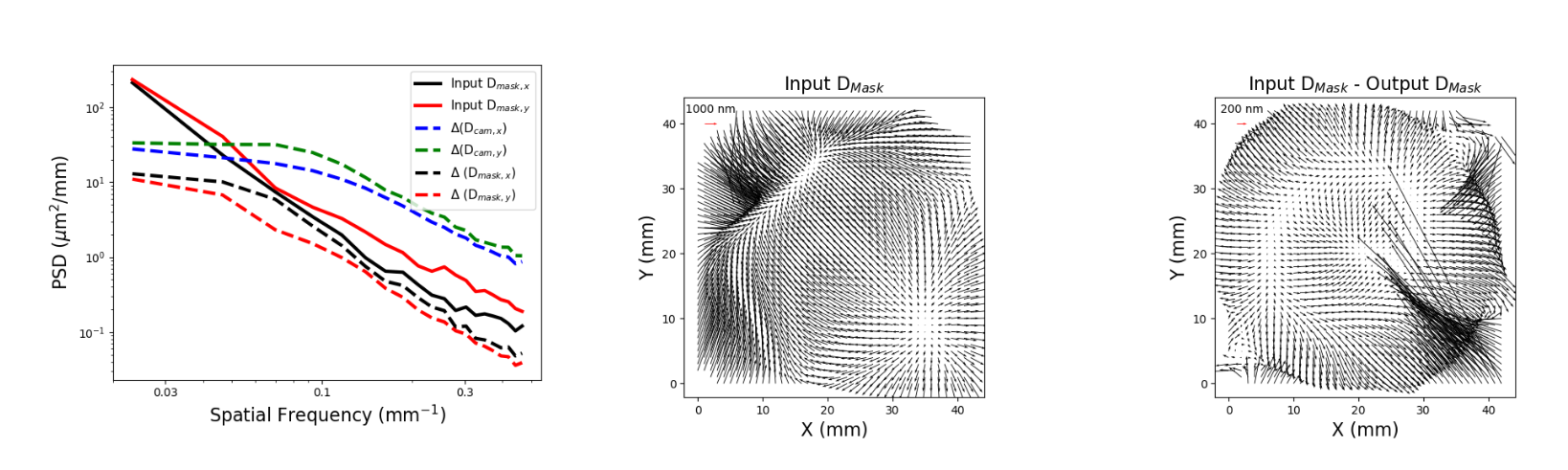}
    \caption{Results from the simulation of the worst case scenario where all of the deviations in the system are due to the mask distortion.  The large recovery error in the mask distortion is a result of a large portion of the mask distortion being mistaken for optical distortion.  The RMS size of the input mask distortion in this simulation is 858 nm, and the size of the recovery error is 198 nm RMS.  We adopt that fractional error of 23\% as the systematic error in the measurement of the mask distortion.}
    \label{fig:sim_all_dist}
\end{figure}

Now we repeat that simulation and analysis; however, instead of using the realistic inputs for both the camera and mask distortion, we instead apply only a mask distortion with an amplitude equal to 858 nm, which is equivalent to the combined mask$+$camera distortion in the previous simulation.  
We derive a suitable model for mask-only distortion by first fitting a $6^{\textrm{th}}$ order Legendre Polynomial to the difference between the measured catalog positions at mask position 0 and a regular square grid.
The resulting input mask-only distortion pattern is plotted in the middle panel of Figure \ref{fig:sim_all_dist}, and it is worth emphasizing that this input does not match expectations for our optical system as the lens is specified to have $\sim$1000 nm of optical distortion over this field.  
However, as a conservative constraint, we use this simulation to estimate the level of systematic error due to misidentifying some of the mask distortion as camera distortion.  
The recovery error is influenced by the outlier points in the lower right corner, so we apply a 5$\sigma$ clip to the residuals before calculating statistics.
This results in a systematic mask distortion recovery error of 200 nm when the input mask distortion had a total size of 858 nm RMS.
Note that there is an even larger mistake in the recovery of the camera distortion, which should be zero, for this simulation, as seen in the PSD of the deviations shown in Figure \ref{fig:sim_all_dist}.

Based on the results from this simulation, we estimate a fractional error of 23\% on the mask distortion, which 
we adopt as the systematic error for the measurement of the mask distortion. 
This corresponds to an additional 10.8 nm of error in the measurement of the mask distortion pattern.

This estimate of the systematic error assumes that the total deviations represent the maximum distortion on the mask or in the camera.
However, there is one case that this approach does not account for.
Specifically, as we chose to only rotate the pinhole mask, radial modes of optical distortion that are centered on our axis of rotation can be modeled as either mask or optical distortion. 
 In the case that there is both a large radial mode of camera distortion centered on the axis of rotation and the opposite mode in the mask distortion they would cancel and not appear in the total deviations, which would lead to an underestimation of the systematic error.
 An ideal calibration unit should have both a rotation and translation stage in order to correctly constrain all modes of distortion.

\subsection{Accuracy of the Self-calibration Method}
\label{sim:error_self_cal}

\begin{figure}
    \centering
    \includegraphics[width=0.5\textwidth]{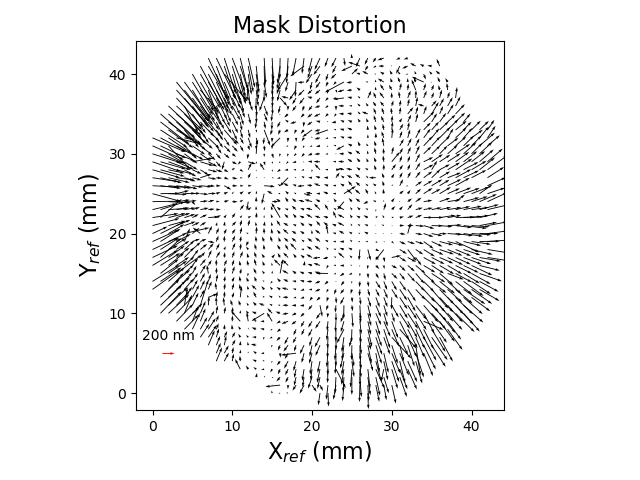}
    \caption{Input mask distortion for the simulations summarized in Table \ref{tab:simulation}.  There is a total of 150 nm RMS of deviations from a square pattern in this pattern.}
    \label{fig:sim_mask_dist}
\end{figure}

We can extend this simulation approach to a wider variety of mask dither patterns in the calibration procedure. 
This is useful as it demonstrates a few simple cases of how the effectiveness of the self calibration approach depends sensitively on how the pinhole mask is moved. 
For these simulations we adopt a 150 nm RMS of mask distortion and 858 nm RMS of camera distortion as shown in Figures \ref{fig:measured_dist} and \ref{fig:sim_mask_dist}. Then we simulate a variety of dither patterns of the mask.
Here we rotate the mask about the center of the detector and translate the mask in a N$_{step}$ by N$_{step}$ square pattern at each of the N$_{\textrm{rot}}$ rotation angles, such that there are N$_{step}^{2}$ measured catalogs for each rotation position.
The translation step size is the spacing between each one of the translation positions of the mask.
These simulations do not include random measurement errors of any kind.

The results for different mask dither patterns are shown in Table \ref{tab:simulation}, cases 1-5.
Column 7 lists the RMS difference between the input mask distortion and the mask distortion recovered by the full fit, which corresponds to the systematic error in the mask distortion measurement due to the chosen dither pattern of the pinhole mask.
Column 8 lists the residual distortion after the full model is fit.
Simulation cases 3-5 shows that translation-only dithers mis-identify 95\% of the input mask distortion as camera distortion.

This degeneracy between mask distortion and camera distortion occurs when the scale between each data set is allowed to vary, which causes aliasing of the mask distortion into camera distortion space.
Appendix 1 demonstrates this effect in detail using a 1D example.
In principle, if the scale is known to be very stable, then it can be fixed in the model, breaking this degeneracy. 
To test the effect of scale variation in the system when scale variation is not allowed in the model, we run a final simulation.
We use the same inputs as simulation case 3, except we add in scale variation for each of the 9 measured catalogs drawn from a normal distribution centered on a scale of 1.0 with a standard deviation of $10^{-5}$.
When we fit this simulation with a model that assumes a fixed scale (allowing for rotation and translation for alignment), we find mask distortion recovery errors and model fit residuals ranging from 20-48 nm RMS in 10 trials.
In contrast, we have an error of only 4.8 nm RMS when there is no input scale variation and a fixed-scale model is used. 
We conclude that using a fixed scale model eliminates the problem of misidentifying mask distortion as camera distortion; however, errors due to the true scale variation in the data can still prevent a precise distortion calibration.
As a final note, scale variations at the $>10^{-5}$ level are commonly seen in astronomical imaging systems and it would take significant additional effort to eliminate them.
Some previous work \cite{Rodeghiero:2019} has claimed that translation-only dithers can fully recover mask distortions, in contrast to our finding. However, these previous simulations only included uncorrelated (i.e. random) mask distortions. 
We simulate this case using 150 nm RMS of uncorrelated mask distortion as the input and find that we can accurately recover the input distortion to 3.8 nm (see Table \ref{tab:simulation} case 6).
This matches the results of previous simulation effort.
However, realistic mask distortions are unlikely to be random as any effect that flexes the pinhole mask (i.e. choice of mount, temperature) and errors in the manufacturing process will produce spatial correlations.
This result emphasizes that future simulations must consider the case of spatially correlated errors in the mask distortion. 

\begin{deluxetable}{ccccccccc}
 \tabletypesize{\footnotesize}
\tablecolumns{8} 
\tablewidth{0pt}
\label{tab:simulation}
\tablecolumns{8}
\tablecaption{Simulation Results}
\tablehead{
\colhead{}
&\colhead{Input}
&\colhead{Input}
&\colhead{}
&\colhead{}
&\colhead{Translation}
&\colhead{Accuracy:}
&\colhead{Precision:} 
&\colhead{} \\
\colhead{Case}
&\colhead{D$_{\textrm{Mask}}$ (nm)}
&\colhead{D$_{\textrm{Cam}}$ (nm)}
&\colhead{N$_{\textrm{rot}}$}
&\colhead{N$_{\textrm{step}}$}
&\colhead{Step size (mm)}
&\colhead{$\Delta$D$_{\textrm{Mask}}$  (nm)\tablenotemark{a}}
&\colhead{Fit Residual (nm)}
&\colhead{Comment}
}
\startdata
1&150 & 858 &12&2&0&108&5 & rotation only \\
2&150 & 858 &12&2&6&6.5&3.1 & translation and rotation \\
3&150 & 858 &1 &3&3&142&0.45 & small translation only \\
4&150 & 858 &1 &6&3&142&0.12 & translation only \\
5&150 & 858 &1 &6&6&142&.09 & translation only \\
6&150 (rnd)\tablenotemark{b} & 858&1&3&3&3.8&0.8&translation only\\
\enddata
\tablecomments{N$_{\textrm{rot}}$ is the number of mask rotation used with them evenly spaced over 360 degrees. N$_{\textrm{step}}$ is the number of translation steps per rotation angle taken along both the x and y axis for each rotation angle.  The step size is the size of each translation step.\\
$^a$ Input - Output mask distortion.\\
$^b$ Ordered mask distortion is used in Simulations 1-5.  Simulation 6 uses uncorrelated mask distortion as the input.
}
\end{deluxetable}

\section{Discussion}
\label{sec:discuss}

Here we have measured the true deviations from the intended pattern for our pinhole mask to be 47.2 nm RMS to a precision of 4.5 nm with an additional systematic error of 10.8 nm. 
This is precise enough that it could offer significant improvement over current distortion measurements for NIRC2 \cite{Service:2016}.
The mask pattern accuracy has important implications for the mechanical design of any distortion calibration unit, and the accuracy of this mask means it could use a static calibration unit.
Additionally, we find that when aligning each catalog with full linear transformations, the mask distortion can easily be mistaken for camera (optical) distortion due to aliasing effects. 
Self-consistent solutions derived from observations that only translate the mask to generate data diversity fail quite badly, with 95\% of the mask distortion misidentified as camera distortion.
An accurate self-consistent solution using our model for both the mask and camera distortion is only achieved when we use full rotational freedom and translations of $\sim$ 15\% of the camera field of view (Simulation 2 in Table \ref{tab:simulation}).
The degree of aliasing depends on the specific distortion model and is particularly sensitive to the need to fit scale changes for each measured catalog.
Real optical systems generally suffer from some scale variation; for example, the lab experiment described in this paper had relative scale variations of order $10^{-5}$ between measurements and the relative error in the plate scale for NIRC2 is $2*10^{-5}$ \cite{Yelda:2010,Service:2016}.

Regardless of the solution to the aliasing problem, the potential degeneracy between mask and camera distortion means that simulations of distortion calibration units must include spatially correlated mask distortion as these modes prove more challenging for the self-calibration method.
It is possible that mask distortion is due to either the manufacturing of the mask or has been introduced via the mount (or both), however, these sources will also be present in realistic distortion calibration units and they influence the requirements for calibration of a given mask.
Instead of self calibrating inside the instrument by moving the mask, it is possible to calibrate the mask to accurately measure the pinhole positions to avoid the systematic errors.
It is worth emphasizing that the choices around mounting the mask are important as it is possible to introduce additional mask distortion by slightly deforming the pinhole mask.

Even though the intrinsic mask distortion is too large for static use in the most demanding instruments, it is still a substantial improvement for some existing distortion solutions.
A distortion solution estimated with only a single position of this pinhole mask would be limited to a systematic precision of 47.2 nm, which corresponds to 23.6 $\mu$as for TMT and 130 $\mu$as for NIRC2. 
That offers a substantial improvement over the current distortion model for NIRC2 instrument at the W.~M.~Keck Observatory that has a total residual distortion of $>$ 1000 $\mu$as, while a measurement using this mask imaged 1:1 would have a systematic error of 130 $\mu$as, and would allow for continued monitoring of the distortion.
It is not intrinsically sufficient for the TMT requirement of 20 nm, but it is possible the pattern errors in the mask could be measured to account for the additional error.
Aside from the mask distortion, the nonlinear optical distortion of the system was measured with a residual of 39 nm over an area of 1788 mm$^{2}$.
One of the sources of this residual is optical distortion that is much higher order than our distortion model, which could be mitigated using a different model.  
This verifies that the approach of using a pinhole mask as an astrometric flat field has the potential to yield excellent results in the most demanding astronomical instruments, assuming that the calibration unit accounts for the mask distortion.  

\section{Acknowledgements}
We would like to acknowledge the NFIRAOS team and NRC-Herzberg for providing the pinhole mask. 
This work was supported by the National Science Foundation under Grant No. 1310706.
The TMT Project gratefully acknowledges the support of the TMT
collaborating institutions.  They are the California Institute of Technology,
the University of California, the National Astronomical Observatory of Japan,
the National Astronomical Observatories of China and their consortium
partners, the Department of Science and Technology of India and their
supported institutes, and the National Research Council of Canada.  This
work was supported as well by the Gordon and Betty Moore Foundation,
the Canada Foundation for Innovation, the Ontario Ministry of Research and
Innovation, the Natural Sciences and Engineering Research Council of
Canada, the British Columbia Knowledge Development Fund, the Association
of Canadian Universities for Research in Astronomy (ACURA), the
Association of Universities for Research in Astronomy (AURA), the U.S.
National Science Foundation, the National Institutes of Natural Sciences of
Japan, and the Department of Atomic Energy of India.

\bibliography{main}
\bibliographystyle{spiejour}

\appendix 
\section{1D Mask Distortion}
\begin{figure}
    \centering
    \includegraphics[width=0.5\textwidth]{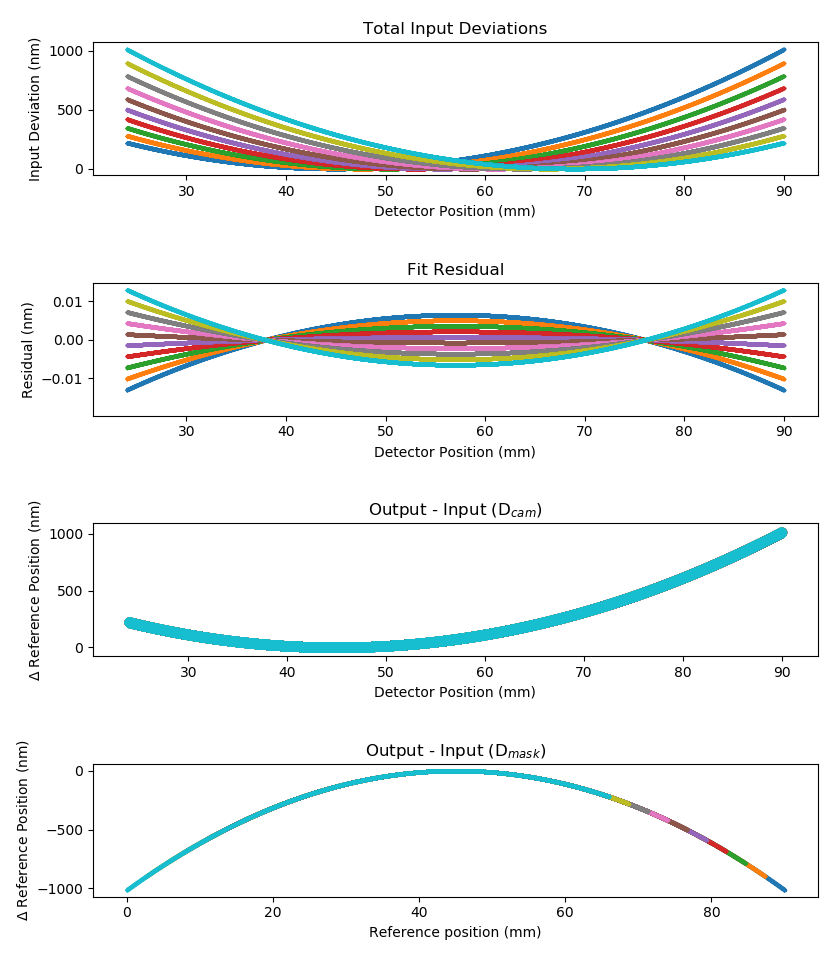}
    \caption{Results from a one dimensional example for the distortion self-calibration problem.   Here each color corresponds to one of the 10 simulated input catalogs.  The detector position is the position on the detector that a source was measured at. For this simulation, there is no input camera distortion and a quadratic term for the mask distortion. We use a model that only includes camera distortion and find that it fits the simulated data with residuals $<$ 0.1 nm.  This results in large recovery errors for the camera distortion (panel 3) and mask distortion (panel 4), which completely fail to accurately describe the system.  This example shows how the model is degenerate, where this a quadratic mode of mask distortion can be fit as either camera distortion or mask distortion.}
    \label{fig:1d_results}
\end{figure}

Here we use a simple 1-dimensional version of the self-calibration problem as an example of how mask distortion can be misidentified as camera distortion.
This situation can be visualized as repeatedly imaging a ruler (reference positions) as it is translated over a camera.
For this simulation we consider a single row of reference positions spaced every 6 $\mu$m with a total length of 90 mm and a detector that is 70 mm long with 300 nm RMS of mask distortion.
To do this, we create an array of evenly spaced reference positions every 6 $\mu$m ($x_{ref}$) and apply a small deviation to each measurement as shown in equation \ref{eq:1d_pat_dev} to generate $x_{ref}'$ which are the true reference positions.
\begin{equation}
    x_{ref}' = x_{ref} + 3*10^{-9}(x_{ref} - 45)^{2}
    \label{eq:1d_pat_dev}
\end{equation}
This set of reference positions is then translated over the camera to 10 times with a step of 2.4 mm each time to create 10 sets of simulated measured data ($x_{cam,n}$), where $n$ ranges from 0 to 9 (equation \ref{eq:1d_sim} and the top panel of figure \ref{fig:1d_results}).
\begin{equation}
    x_{cam,n} = x_{ref}' + 2.4 * n
    \label{eq:1d_sim}
\end{equation}
Camera distortion could be applied here as a function of the measured camera position ($x_{cam,n}$), however, we do not input any camera distortion in this simulation.
An accurate model fit of this simulation will recover 0 nm of optical distortion and 300 nm of mask distortion.

\begin{figure}
    \centering
    \includegraphics[width=0.5\textwidth]{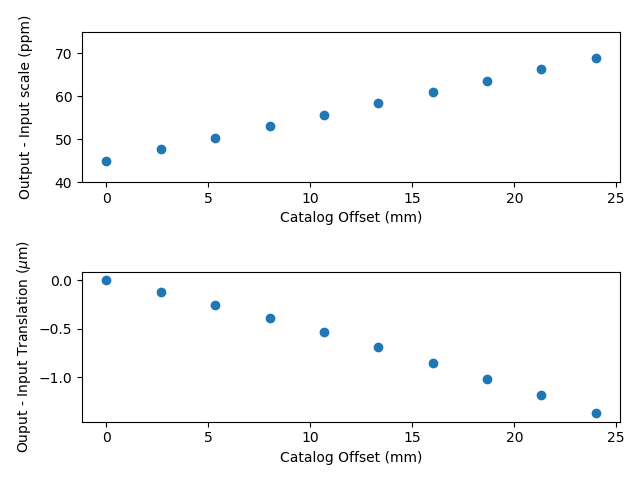}
    \caption{Recovery errors for the linear transformations between each catalog ($a_{n}$ and $b_{n}$ respectively). Using a fixed scale is a way to break the degeneracy between mask and camera distortion, however, this places more stringent requirements on the optical system. 
}
\label{fig:1D_lin_errors}
\end{figure}

We choose to fit these simulated measurements with a model that only includes camera distortion.
For this purpose, we use a Cartesian polynomial up to order 4 as shown in equation \ref{eq:1d_camdist}.
\begin{equation}
    x_{model,n} = x_{cam,n} + c_{1}x_{cam,n}^{2} + c_{2}x_{cam,n}^{3} + c_{3}x_{cam,n}^{4}
    \label{eq:1d_camdist}
\end{equation}
Here $x_{cam}$ is the measured position on the detector, $x_{model}$ is the distortion corrected position for each of the $N$ catalogs, the $c$ coefficients are the camera distortion model.
The model is fit by minimizing $\Delta$ as defined in equation \ref{eq:1d_min}
\begin{equation}
    \Delta = \sum^{N} x_{model,n} - (a_{n} + b_{n}x_{ref})
    \label{eq:1d_min}
\end{equation}
Here $x_{model,N}$ are the distortion corrected measured positions, $x_{ref}$ are a set of evenly spaced input coordinates,the $a$ and $b$ coefficients are the linear transformation parameters for each of the $N$ catalogs.
These linear transformations account for the scale variation in a real system ($a_{n}$) and the unknown amount of translation from an imprecise stage ($b_{n}$).
Note that this approach is comparable to a 2D self calibration problem which allows for full linear (6 parameter) transformations between each catalog.
As the only input deviations are mask distortion and our model only includes camera distortion, we expect that this model should {\em not} precisely describe this simulated system.
The second panel of Figure \ref{fig:1d_results} demonstrates that the camera distortion model can fit the input deviations with  a fit residual of less than 0.002 \% of the input distortions.
This is the worst case scenario, as the errors in the reference positions have been misinterpreted as camera distortion in the system.
One solution to this issue is to rotate the reference positions in order to generate greater data diversity.
The other possible solution is to fix the scale in the alignment transformation.  
As Figure \ref{fig:1D_lin_errors} shows, the best fit model has errors larger  than 1 $\mu$m in the recovery of the translation of the reference positions ($a_{n}$) as well as scale errors as large as $7 *10^{-5}$.
These inaccuracies point to the another way of solving the degeneracy; if we assume that the magnification of the optical system is the same for each data and only solve for the translation offset between each catalog we accurately recover the mask and camera distortion.
This is consistent with the 2D simulation results in the main text.
\end{document}